\documentclass[twocolumn,08 pt,amsmath,amssymb,aps,fleqn]{revtex4-1}
\usepackage[hidelinks]{hyperref}
\usepackage[english]{babel}
\usepackage{mathtools}
\usepackage{tabularx}
\usepackage{multirow}
\usepackage{comment}
\usepackage{graphicx}
\usepackage{color}
\usepackage{float}
\usepackage{soul,cancel}
\begin{document}

\preprint{APS/123-QED}
\title{Impurity-induced Friedel oscillations and Wigner crystallization in Luttinger liquids via Unified Field Bosonization Technique}
\author{Soundarya P$^{ 1}$, Venkata Suryanarayana M$^{ 1}$ and Joy Prakash Das$^{ 2*}$}
\affiliation{
	$^{\it 1}$Department of Physics, National Institute of Technology Tiruchirappalli, Tamil Nadu - 620015, India\\
	$^{\it 2}$Department of Physics, Assam Engineering College, Guwahati, Assam - 781013, India
		   }
\email{jpdas100@gmail.com}
\begin{abstract}
\begin{center}\bfseries Abstract\end{center}
We present an analytical study of impurity-induced density correlations in interacting one-dimensional Luttinger liquids using the Unified Field Bosonization Technique (UFBT). Unlike conventional approaches that treat impurities perturbatively, UFBT incorporates localized impurities exactly at the free-fermion level while treating interactions non-perturbatively. Closed analytical expressions for fermionic Green functions and density-density correlation functions are obtained for strongly inhomogeneous systems, enabling a unified description of the impurity-induced $2k_F$ Friedel oscillations and the $4k_F$ Wigner-crystal like correlations. The corresponding scaling behaviour is derived analytically, revealing a crossover in the dominant long-distance correlations at $g=1/3$: Friedel correlations dominate for $g>1/3$, whereas Wigner-crystal like correlations become dominant for $g<1/3$, corresponding to the regime of sufficiently strong repulsive interactions. A reflection-coupled harmonic analysis is developed to account for the left-right mixing induced by impurity backscattering and to consistently reproduce the resulting density correlations. The spinless limit is also obtained as a direct reduction of the spinful formulation. These results provide an analytically controlled framework for describing impurity-induced and interaction-driven density correlations in strongly inhomogeneous one-dimensional quantum systems.

\end{abstract}
\maketitle
\section{Introduction}

The presence of a localized scatterer can profoundly modify the particle density of a one-dimensional quantum system, producing oscillations that extend away from the scattering region. For noninteracting fermions, these oscillations originate from the interference between incident and reflected waves and constitute the familiar Friedel oscillations. First predicted by Jacques Friedel in the context of metallic alloys \cite{friedel1952xiv, friedel1958metallic}, Friedel oscillations have subsequently been observed through scanning tunneling microscopy and related spectroscopic techniques in a variety of low-dimensional and correlated materials, including noble-metal surfaces \cite{crommie1993imaging,hasegawa1993direct}, graphene \cite{bena2016friedel,dutreix2016friedel}, altermagnets \cite{chen2024impurity}, and superconducting systems \cite{stosiek2022friedel}. In one dimension, the oscillatory component is characterized by the wave vector $2k_F$, while its long-distance envelope is governed by the low-energy properties of the underlying fermionic system \cite{das2020friedel}. This physics is relevant to semiconductor quantum wires \cite{auslaender2002tunneling}, carbon nanotubes \cite{ishii2003direct}, and quantum Hall edge states \cite{moon1993resonant}. Recent studies have further shown that quantum geometry can give rise to distinct oscillation periods and decay lengths in flat-band systems \cite{ma2026quantum}, while spin-dependent Fermi-surface anisotropy can produce direction-dependent Friedel-oscillation periods in altermagnetic systems \cite{chen2024impurity}. These developments emphasize the sensitivity of Friedel oscillations to the microscopic and collective properties of the host system.

In one dimension, electron-electron interactions strongly influence the long-distance behavior through collective fluctuations, making the Luttinger-liquid framework a natural description of impurity-induced density modulations \cite{haldane1981luttinger,giamarchi2004quantum}. The impurity problem in a Luttinger liquid differs qualitatively from that in a conventional Fermi liquid: a localized impurity couples right- and left-moving modes, while interactions modify the scaling properties of the resulting density response. The renormalization-group analysis of Kane and Fisher established that a localized barrier constitutes a nontrivial perturbation, whose low-energy behavior depends on the interaction strength \cite{kane1992transport,aristov2008transport,aristov2009conductance}. For repulsive interactions, an arbitrarily weak impurity is a relevant perturbation and flows toward the strong-coupling limit, where the system approaches an effectively disconnected configuration. Consequently, the density modulation generated by an impurity is not simply the non-interacting Friedel pattern with a modified amplitude; interactions alter its long-distance power-law behavior.

Several analytical approaches have been developed to study impurity-induced Friedel oscillations in interacting one-dimensional systems. Egger and Grabert obtained interaction-dependent spatial decay laws for arbitrary impurity strength \cite{egger1995friedel,egger1996friedel}, while related analytical treatments examined the impurity-induced density response and its spatial dependence \cite{grishin2004functional}. Path-integral formulations have been employed for more general forms of electron-electron interactions \cite{fernandez2001friedel}, and boundary conformal field theory has provided a systematic description of the asymptotic scaling behavior near an impurity. Exact results are also available at particular interaction parameters; notably, the case corresponding to a Luttinger parameter $g=1/2$ permits an exact treatment for arbitrary impurity coupling \cite{leclair1996exact}. Numerical approaches, including density-matrix renormalization group methods, provide complementary studies of impurity problems in interacting one-dimensional systems \cite{white1992density,schollwock2005density,qin1996impurity,hamamoto2008numerical}.

At sufficiently low densities and strong repulsive interactions, one-dimensional electron systems can develop pronounced Wigner-crystal-like correlations, arising from the dominance of Coulomb repulsion over kinetic energy \cite{meyer2009wigner,ziani2020short}. Such behavior has been investigated in quantum wires and finite few-electron systems, where spatial density peaks associated with crystalline charge localization can emerge despite the absence of true long-range order in one dimension \cite{meyer2009wigner,vu2020one}. Wigner correlations have also been studied in Luttinger-liquid quantum dots and confined carbon nanotubes, with correlation functions and the spatial structure of the electron density providing signatures of the correlated state \cite{gambetta2014correlation,sarkany2017wigner}. Importantly, in finite one-dimensional Luttinger systems, Friedel oscillations and Wigner correlations can exhibit the same spatial wavelength, making their distinction nontrivial from the density profile alone \cite{gambetta2014correlation}.

Despite these advances, obtaining closed analytical expressions for density correlations in the presence of impurities of arbitrary strength and interactions remains considerably more involved, particularly for spatially extended and strongly inhomogeneous impurity configurations \cite{das2018quantum,das2019nonchiral,danny2020density}. The Unified Field Bosonization Technique (UFBT), developed in our previous work, incorporates such inhomogeneity directly into the Fermi--Bose correspondence before the interaction-induced collective dynamics are introduced \cite{das2026unified}. This formulation provides direct access to $N$-point density correlation functions, offering a simple route to the density response in the presence of impurities without requiring a subsequent renormalization-group treatment of the impurity.

In the present work, this framework is applied to impurity-induced density oscillations in interacting and strongly inhomogeneous one-dimensional Luttinger liquids. The $2k_F$ Friedel oscillations are obtained from the corresponding two-point density-density correlation function, while higher-order correlations associated with Wigner-crystal-like behavior are investigated through the four-point density correlation function. Closed analytical expressions are obtained for arbitrary impurity and interaction strengths, allowing the spatial dependence and power-law behavior of the different contributions to be examined on the same footing. The Friedel oscillations and their envelope, the rapidly oscillating part of the density-density correlation function, and the Wigner-crystal-like correlations are analyzed and compared. In addition, an alternative harmonic analysis appropriate for strongly inhomogeneous systems is developed, in which both $\rho(y)$ and $\rho(-y)$ enter the density representation. This provides a unified analytical description of the density correlations and their associated oscillatory behavior in the presence of strong impurity scattering and electron-electron interactions.


\section{System Description and Methodology}

We consider an interacting one-dimensional electron system containing a localized static impurity structure centered at $x=0$. The system is described by the Hamiltonian.
\begin{equation}
\begin{aligned}
H ={}& \int_{-\infty}^{\infty} dx\, \psi^\dagger(x) \left[ -\frac{1}{2m}\partial_x^2+V(x) \right] \psi(x)\\
&+\frac{1}{2}\int_{-\infty}^{\infty}dx \int_{-\infty}^{\infty}dx'\,v(x-x')\rho(x)\rho(x'),
\end{aligned}
\label{hamiltonian}
\end{equation}
where  $V(x)$ denotes the localized impurity potential, and $v(x - x') = \frac{1}{L} \sum_q v_q \exp[-iq(x - x')]$ represents the electron-electron interaction. As in the previous UFBT formulation, the interaction is restricted to the forward-scattering sector. For the short-range interaction considered here, the Fourier components may be written as
\begin{equation}
\hspace{2 cm}
v_q =
\begin{cases}
v_0, & |q|<\Lambda,\\
0, & |q|>\Lambda,
\end{cases}
\label{interaction}
\end{equation}
where $\Lambda$ ($\ll k_F$) is a momentum cutoff.
The impurity potential is assumed to be confined to a finite region around the origin and may contain an arbitrary combination of localized barriers and wells. Rather than specifying the detailed form of $V(x)$, the scattering properties of the impurity region are characterized by its reflection and transmission amplitudes, denoted by $R$ and $T$, respectively. Thus, the microscopic structure of the impurity enters the low-energy correlation functions through its single-particle scattering data. The normalization condition $|R|^2+|T|^2=1$ holds for the corresponding elastic scattering problem.

At low energies, the fermionic field is separated into contributions associated with the two Fermi points,
\begin{equation}
\psi(x)=e^{ik_Fx}\psi_R(x)+e^{-ik_Fx}\psi_L(x).
\label{fermiondecomposition}
\end{equation}
The density \big($\rho(x)=\psi^\dagger(x)\psi(x)$\big) consequently contains a smooth contribution as well as terms oscillating with wave vector $2k_F$. 
In the RPA sense the density may be harmonically analysed as 
\begin{equation}
\rho(x,\sigma,t)=\rho_s(x,\sigma,t)+e^{2ik_Fx}\rho_f(x,\sigma,t)+e^{-2ik_Fx}\rho_f^*(x,\sigma,t)
\label{rho2kf}
\end{equation}
In particular, the component responsible for the Friedel oscillation can be written as
\begin{equation}
\rho_{2k_F}(x)=e^{2ik_Fx}\psi_L^\dagger(x)\psi_R(x)+e^{-2ik_Fx}\psi_R^\dagger(x)\psi_L(x)
\label{rho2kf}
\end{equation}
The impurity-induced density modulation is therefore obtained from the mixed-chirality fermionic correlation functions associated with the two terms in equation (\ref{rho2kf}).
The calculation employs the Unified Field Bosonization Technique developed recently \cite{das2026unified}. The defining prescription of UFBT is to construct the bosonized fermionic field using density and current fields associated with both sides of the inhomogeneous region. In the notation used here, the corresponding prescription is
\begin{equation} 
\psi_\nu(x,\sigma,t) \sim \exp\left[\frac{i}{\sqrt{2}}\left\{\theta_\nu(x,\sigma,t)+\theta_\nu(-x,\sigma,t)\right\}\right]
\label{ufbt} \end{equation}
with the local phase given by the formula,
\small
\begin{equation}
\begin{aligned}
\theta_{\nu}(x,\sigma,t) = \pi \int^{x}_{sgn(x)\infty}& dy \bigg( \nu  \mbox{  } \rho_s(y,\sigma,t) \\
&- \int^{y}_{sgn(y)\infty} dy^{'} \mbox{ }\partial_{v_F t }  \mbox{ }\rho_s(y^{'},\sigma,t) \bigg)
\end{aligned}
\end{equation}\normalsize
where $\nu=+1$ and $\nu=-1$ denote the right and left moving branches, respectively. The symbol $\sim$ indicates that equation (\ref{ufbt}) is a prescription for constructing correlation functions rather than an exact operator identity. Any system-dependent coefficients associated with the fermionic fields are determined independently from the corresponding non-interacting correlation functions.	

The density correlations entering equation (\ref{ufbt}) are first obtained for the non-interacting inhomogeneous system, with the reflection and transmission amplitudes retaining the information about the impurity configuration. The forward-scattering interaction is then incorporated through the corresponding interaction-dressed density-density correlation functions. For the short-range interaction in equation (\ref{interaction}), the charge and spin sectors are characterized by the velocities
\begin{equation}
v_h = \sqrt{ v_F^2+\frac{2v_Fv_0}{\pi}}, \qquad v_n=v_F ,
\label{velocities}
\end{equation}
where $v_h$ and $v_n$ denote the holon and spinon velocities, respectively. The two sectors remain decoupled within the present forward-scattering description.
The interacting fermionic correlation functions are subsequently constructed by substituting the interaction-modified density correlations into the UFBT prescription. For the present problem, only the mixed-chirality correlator required by equation (\ref{rho2kf}) is needed. The explicit form of these quantities is obtained from the UFBT correlation functions in the following section. The above construction retains the impurity scattering exactly at the single-particle level while incorporating the forward-scattering interaction through the interacting density correlations. It therefore provides the starting point for deriving the Friedel oscillations without introducing a perturbative expansion in either the impurity strength or the interaction strength.

\section{ Results and Discussion}
\subsection{Friedel Oscillations and its envelope}

Within the low-energy description, the fermionic density is decomposed into slowly varying and oscillatory contributions. The latter contains the $2k_F$ component responsible for impurity-induced Friedel oscillations,
\small
\begin{equation}
\rho(x,\sigma,t) = \rho_s(x,\sigma,t) + e^{2ik_Fx}\rho_f(x,\sigma,t) + e^{-2ik_Fx}\rho_f^*(x,\sigma,t)
\label{densitydecomposition}
\end{equation}
\normalsize
where $\rho_s=\psi^{\dagger}_R\psi_R+\psi^{\dagger}_L\psi_L$ denotes the slowly varying density, while $\rho_f=\psi^{\dagger}_L\psi_R$ and $\rho_f^*=\psi^{\dagger}_R\psi_L$ represent the slowly varying envelopes associated with the $2k_F$ and $-2k_F$ harmonics, respectively.

The two-point function $\langle \psi_R \psi_L^{\dagger} \rangle$ obtained using the UFBT \cite{das2026unified} is given by \scriptsize
\begin{equation*}
\begin{aligned}
\Big\langle T&\mbox{ }\psi_{R}(x_1,\sigma_1,t_1)\psi_{L}^{\dagger}(x_2,\sigma_2,t_2)\Big\rangle \sim
\frac{\delta_{\sigma_1,\sigma_2}}{[(x_1-x_2)^2 -v_h^2 \tau_{12}^2]^{X}} \\
&\times \frac{(4x_1x_2)^{X}}{ (x_1+x_2 -v_h \tau_{12})^{P} (-x_1-x_2 -v_h \tau_{12})^{Q} (x_1+x_2 -v_F \tau_{12})^{0.5}}\
\end{aligned}
\end{equation*}
\normalsize
where the correlation function exponents are given by\small
\begin{equation}
P = \frac{(v_h+v_F)^2}{8v_hv_F}
\ ;\quad
Q = \frac{(v_h-v_F)^2}{8v_hv_F}
\ ;\quad
X = \frac{v_h^2-v_F^2}{8v_hv_F}.
\label{exponents}
\end{equation}
\normalsize

To extract the spatial dependence of the oscillatory contribution, we consider the equal-time limit and take $x_1=x$ and $x_2=x+\epsilon$, where $\epsilon$ serves as a short-distance regulator. Thus, with $t_1=t_2$ and consequently $\tau_{12}=t_1-t_2=0$, one obtains
\begin{equation*}
\begin{aligned}
\Big\langle  {\rho}_f(x)  \Big\rangle  \sim  \frac{[4x^2]^X}{\epsilon^{2X}[2x]^P[-2x]^Q[2x]^{0.5}}
\sim \mbox{ } x^{2X-P-Q-\frac{1}{2}}
\end{aligned}
\end{equation*}
Using equations (\ref{exponents})$, 2X-P-Q = -\frac{v_F}{2v_h}$. Now $\frac{v_F}{v_h}=g$ is the well known Luttinger liquid interaction parameter, and therefore
\begin{equation}
\begin{aligned}
\langle  {\rho}_f(x) \rangle  \mbox{ }\sim \mbox{ } (x)^{-(1+g)/2}
\label{FriedelExp}
\end{aligned}
\end{equation}
This reproduces the relation obtained by Egger et al. \cite{egger1995friedel} for the envelope of the oscillatory part of the density, which is accompanied by the characteristic $2k_F$ spatial oscillation.

\begin{figure}[h]
\begin{center}
\includegraphics[scale=0.35]{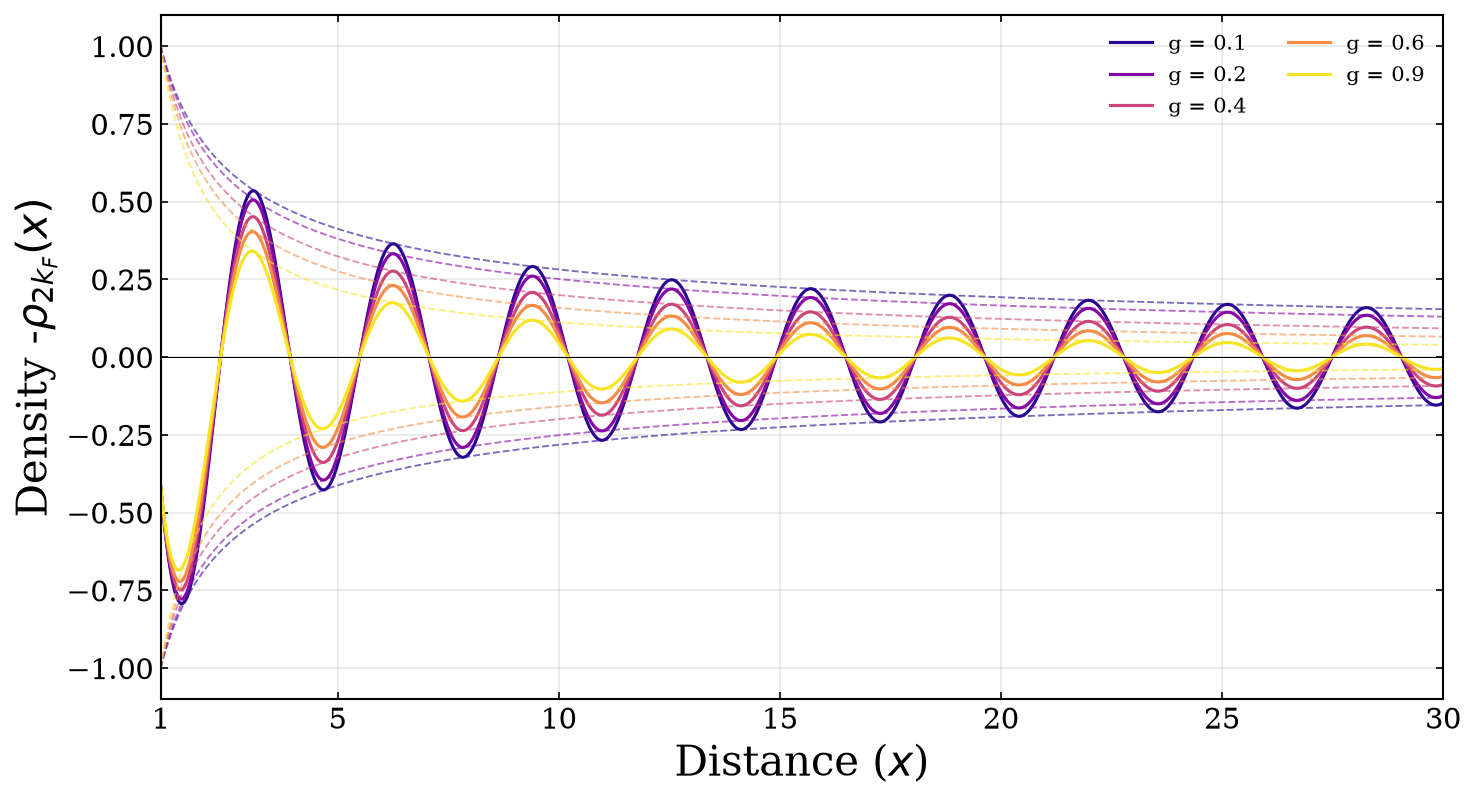}
\end{center}
\caption{Friedel oscillations and their asymptotic envelopes for different strengths of mutual interactions.}
\label{FriedelPlot}
\end{figure}
In Figure (\ref{FriedelPlot}), the impurity-induced $2k_F$ density component, $<\rho_{2k_F}(x)>$, is shown as a function of distance $x$ for several values of the Luttinger parameter $g$. The oscillatory contribution is accompanied by its corresponding power-law envelope, illustrating the interaction-dependent decay of the Friedel oscillations. Increasing $g$, corresponding to weaker electron-electron interactions, leads to a faster power-law decay of the oscillation amplitude. The plots are evaluated for \(x>1\), focusing on the distance range where the asymptotic spatial behavior becomes relevant.

\subsection{Rapidly Oscillating Density-Density Correlations}

The \(2k_F\) Friedel oscillations discussed above characterize the density modulation induced by impurity scattering. The same UFBT framework also permits the direct evaluation of density--density correlations and, in particular, the rapidly oscillating part associated with the interference between right- and left-moving fermionic fields. Such correlations provide information on the spatial structure of density fluctuations beyond the one-point Friedel response.

Using the Fermi--Bose correspondence given in equation (\ref{ufbt}), general \(N\)-point fermionic correlation functions can be evaluated within UFBT \cite{das2026unified}. The density--density correlation relevant here follows as a special case of the four-point fermionic correlation function. For compactness, the notation $X_i\equiv(x_i,\sigma_i,t_i)$ is used. One has
\begin{equation}
\begin{aligned}
\Big\langle T\mbox{ }\psi_{R}(X_1)&\psi_{L}^{\dagger}(X_2)\psi_{R}(X_3)\psi_{L}^{\dagger}(X_4)\Big\rangle \\
\sim &\Big\langle e^{i\Theta_{R}(X_1)} e^{-i\Theta_{L}(X_2)} e^{i\Theta_{R}(X_3)} e^{-i\Theta_{L}(X_4)}\Big\rangle
\end{aligned}
\label{fourpoint}
\end{equation}
where
\begin{equation}
\Theta_{\nu}(X_i)
=\frac{1}{\sqrt{2}}
\left[
\theta_\nu(x_i,\sigma_i,t_i)
+\theta_\nu(-x_i,\sigma_i,t_i)
\right].
\label{bigtheta}
\end{equation}
The exponential average can be evaluated using the Baker--Campbell--Hausdorff (BCH) relation, allowing it to be expressed in terms of the corresponding two-point correlations of the bosonic fields:
\small
\begin{equation}
\begin{aligned}
\Big\langle &e^{i\Theta_{R}(X_1)} e^{-i\Theta_{L}(X_2)} e^{i\Theta_{R}(X_3)} e^{-i\Theta_{L}(X_4)}\Big\rangle\\
= & e^{-\frac{1}{2}<\Theta_{R}(X_1)^2>}e^{-\frac{1}{2}<\Theta_{L}(X_2)^2>}e^{-\frac{1}{2}<\Theta_{R}(X_3)^2>}e^{-\frac{1}{2}<\Theta_{L}(X_4)^2>}\\
&e^{<\Theta_{R}(X_1)\Theta_{L}(X_2)>}e^{<\Theta_{R}(X_1)\Theta_{L}(X_4)>}e^{<\Theta_{L}(X_2)\Theta_{R}(X_3)>}\\
&e^{<\Theta_{R}(X_3)\Theta_{L}(X_4)>}e^{-<\Theta_{R}(X_1)\Theta_{R}(X_3)>}e^{-<\Theta_{L}(X_2)\Theta_{L}(X_4)>}
\end{aligned}
\label{fourpointbch}
\end{equation}
\normalsize
To obtain the density--density correlation function, the fermionic operators are subsequently brought together using point splitting. For the first density operator, the coordinates are assigned as $x_1=x+\epsilon$ and $x_2=x$, with $t_1=t_2=t$. Similarly, for the second density operator, one takes $x_3=x'$ and $x_4=x'+\epsilon$, with $t_3=t_4=t'$. The regulator $\epsilon$ is retained throughout the calculation and is taken to zero only after all relevant two-point correlations have been combined. The spin indices are taken to be equal, \(\sigma=\sigma'\). Thus,
\small
\begin{equation*}
\begin{aligned}
\langle \rho_f(x,&\sigma, t)\rho_f(x',\sigma, t')\rangle\\
=&\lim_{\epsilon\rightarrow0}
\Big\langle T\mbox{ }\psi_{R}(x,t)\psi_{L}^{\dagger}(x+\epsilon,t)\psi_{R}(x',t')\psi_{L}^{\dagger}(x'+\epsilon,t')\Big\rangle 
\end{aligned}
\end{equation*}\normalsize
Substitution of these coordinates into equations (\ref{fourpoint}) and (\ref{fourpointbch}) gives, for the explicitly spatially dependent part,
\footnotesize
\begin{equation}
\begin{aligned}
\Big\langle T&\mbox{  } {\rho}_f(x_1,\sigma,t_1)  {\rho}_f(x_2,\sigma,t_2)\Big\rangle  \sim  \\
&\left(\frac{[(x_1-x_2)-v_h(t_1-t_2)][-(x_1-x_2)-v_h(t_1-t_2)]}{[(x_1+x_2)-v_h(t_1-t_2)][-(x_1+x_2)-v_h(t_1-t_2)]}\right)^{\frac{g}{2}}\\
&\left(\frac{[(x_1-x_2)-v_F(t_1-t_2)][-(x_1-x_2)-v_F(t_1-t_2)]}{[(x_1+x_2)-v_F(t_1-t_2)][-(x_1+x_2)-v_F(t_1-t_2)]}\right)^{\frac{1}{2}}
\label{rhofrhof}
\end{aligned}
\end{equation}
\normalsize
where time-independent spatial factors have been omitted for compactness. Here $g=\frac{v_F}{v_h}$ is the previously discussed Luttinger liquid parameter.
The above expression shows that the density--density correlation is governed by two distinct spatial scales. The factors involving $x_1-x_2$ describe the correlation between density fluctuations at the two observation points, while the appearance of $x_1+x_2$ reflects the influence of the spatial inhomogeneity through the symmetrized UFBT fields. The two factors also involve the velocities $v_h$ and $v_F$, thereby retaining the effects of interactions and the underlying fermionic dynamics, respectively. In the homogeneous limit, the dependence on $x_1+x_2$ associated with the inhomogeneity is removed, and the correlation reduces to the corresponding translationally invariant form.

The rapidly varying part of this density--density correlation contains the familiar $2k_F$ spatial harmonic discussed above, with its amplitude and algebraic envelope determined by the correlation factors in the above expression. Thus, the present calculation establishes the density--density correlations within the same UFBT framework and provides the starting point for examining higher harmonics of the density fluctuations.

A natural extension is to consider correlations between higher-order oscillatory density components. In particular, the $4k_F$ sector is closely connected with the enhanced density correlations that emerge in the strongly interacting regime and with the quasi-long-range Wigner-crystal-like behavior of one-dimensional quantum systems. This requires a separate higher-order density correlation function, which is considered next.
 
\subsection{Wigner Crystal like correlations}
The $4k_F$ harmonic can be generated by the product of the $2k_F$ bilinears associated with the two spin components. In particular, the contribution involving opposite spins can be written as
\begin{equation}
\rho_{4k_F}(x)
\sim
e^{4ik_Fx}
\psi^\dagger_{L\uparrow}(x)\psi_{R\uparrow}(x)
\psi^\dagger_{L\downarrow}(x)\psi_{R\downarrow}(x)
+\mathrm{h.c.}
\end{equation}
Thus, unlike the $2k_F$ contribution, which involves a single right--left fermionic bilinear, the $4k_F$ harmonic involves a product of two such bilinears, one for each spin component. Within the present UFBT treatment, the corresponding contribution is therefore expressed in terms of a four-fermion structure involving the two spin components. In the bosonized representation, this structure takes the form
\begin{equation}
\begin{aligned}
\rho_{4k_F}(x)\sim
\Big\langle
&e^{i\Theta_{R\uparrow}(x)}
e^{-i\Theta_{L\uparrow}(x)}
e^{i\Theta_{R\downarrow}(x)}
e^{-i\Theta_{L\downarrow}(x)}
\Big\rangle 
\label{wigner}
\end{aligned}
\end{equation}\normalsize
where $\Theta_{\nu}(x)$ is defined in equation (\ref{bigtheta}). The expression for the envelope of the $4k_F$ density oscillation term given in equation (\ref{wigner}) is a special case of the four point function $\langle \psi^\dagger_L \psi_R \psi^\dagger_L \psi_R \rangle$ with equal positions and times whereas the spin of the first two points are opposite to that of the second two points. Using a similar treatment as in the previous section, using necessary point splitting and the BCH expansion, the Wigner Crystal term is obtained as follows 
\begin{equation}
\begin{aligned}
\rho_{4k_F}(x)\sim
x^{2(4X-2P-2Q)}
\label{wigner2}
\end{aligned}
\end{equation}
Using equations (\ref{exponents})$,4X-2P-2Q = -\frac{v_F}{v_h}=-g$, the Luttinger Liquid interaction parameter.
\begin{equation}
\begin{aligned}
\rho_{4k_F}(x)\sim
x^{-2g}
\label{wigner3}
\end{aligned}
\end{equation}
 It is important to note that the spinon exponent is absent from the present $4k_F$ correlation. In the interacting system, the slowly varying part of the density--density correlation can be decomposed into holon and spinon contributions, with the relative sign of the spinon term determined by the product of the spin indices of the two density operators \cite{das2026unified}. For equal spins, this product is positive, whereas for opposite spins it is negative. In the present four-point correlation, the up and down spin contributions occur symmetrically. Consequently, the spinon correlation functions cancel pairwise, leaving only the holon contribution.
The resulting $4k_F$ correlation therefore contains no non-trivial spinon exponent. This is in contrast to the $2k_F$ correlation, where the spin sector contributes the trivial exponent $1/2$, giving the characteristic factor $x^{-(g+1)/2}$. Thus, the decay of the $4k_F$ correlation is governed entirely by the charge sector, leading to the exponent $2g$.\\

\begin{figure}[h]
\begin{center}
\includegraphics[scale=0.35]{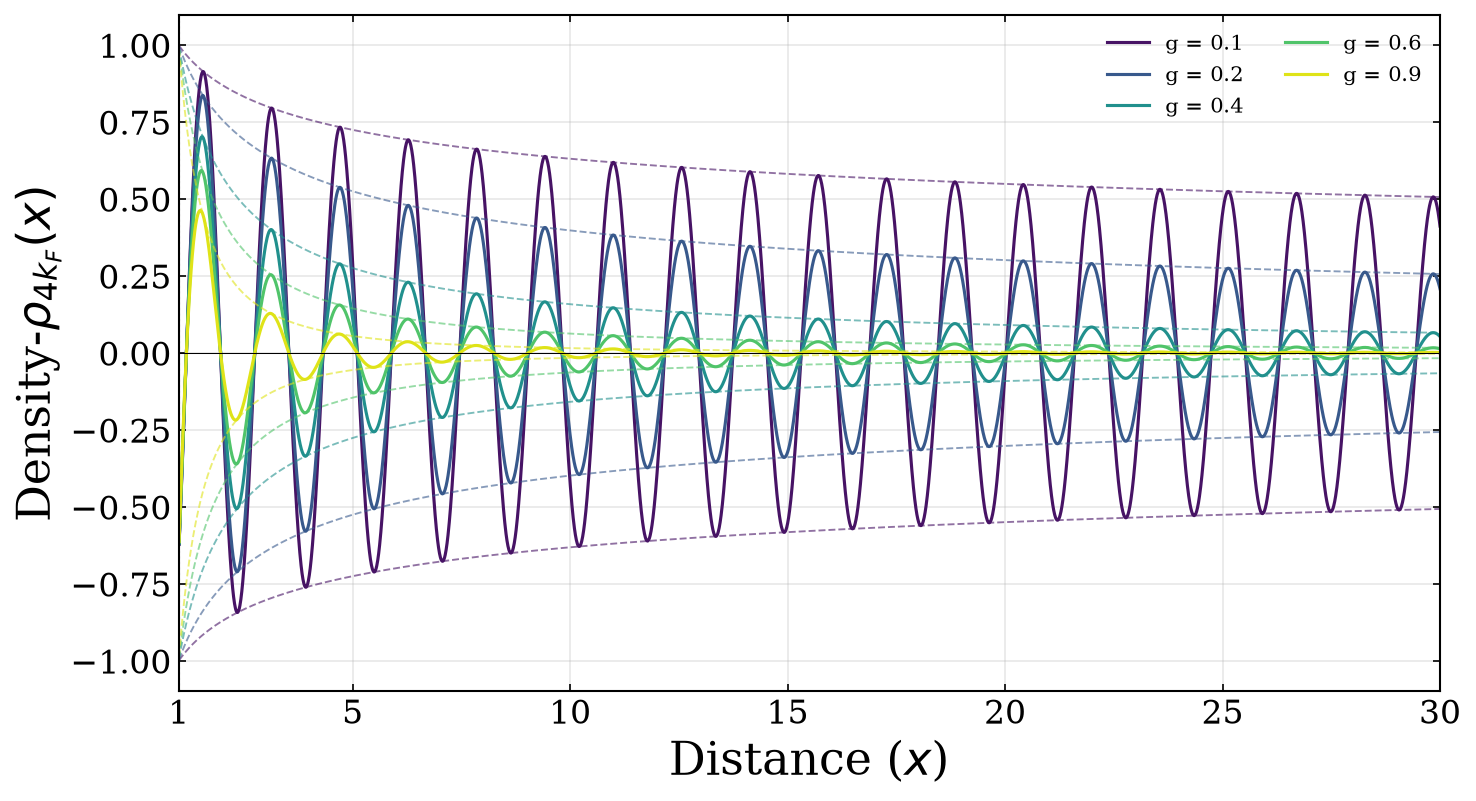}
\end{center}
\caption{Wigner Crystal like density oscillations and their asymptotic envelopes for different strengths of mutual interactions.}
\label{WignerPlot}
\end{figure}
In Figure (\ref{WignerPlot}), the \(4k_F\) component, \(<\rho_{4k_F}(x)>\), is shown for several values of the Luttinger parameter \(g\), together with the corresponding power-law envelopes. The increasing relative persistence of the \(4k_F\) component with decreasing \(g\) reflects the enhanced spatial correlations in the strongly interacting regime.  Increasing $g$, corresponding to weaker electron-electron interactions, leads to a faster power-law decay of the oscillation amplitude. The plots are evaluated for \(x>1\), focusing on the distance range where the asymptotic spatial behavior becomes relevant.

\subsection{Crossover between Friedel and Wigner-crystal correlations}

The competition between the impurity-induced Friedel oscillations and the Wigner-crystal like correlations can be characterized by their respective long-distance scaling behaviours. The Friedel contribution decays as
\begin{equation}
\langle\rho_{2k_F}(x)\rangle\sim x^{-(g+1)/2},
\end{equation}
whereas the Wigner-crystal contribution exhibits the asymptotic behaviour
\begin{equation}
\langle\rho_{4k_F}(x)\rangle\sim x^{-2g}.
\end{equation}
The crossover between the two regimes is therefore determined by
\begin{equation}
\frac{g+1}{2}=2g,
\end{equation}
giving
\begin{equation}
g_c=\frac{1}{3}.
\end{equation}
For $g>1/3$, the $2k_F$ Friedel contribution decays more slowly and consequently dominates the long-distance density profile. In contrast, for $g<1/3$, corresponding to sufficiently strong repulsive interactions, the $4k_F$ contribution decays more slowly and becomes the dominant long-distance correlation.

This crossover has a direct physical interpretation. Friedel oscillations originate from the response of the electron density to the impurity-induced breaking of translational invariance, whereas the $4k_F$ Wigner-crystal-like correlations reflect the enhanced tendency toward spatial ordering arising from strong electron-electron repulsion. As the repulsive interaction becomes sufficiently strong, the relative contribution of the interaction-driven $4k_F$ component increases compared with the impurity-induced $2k_F$ modulation. Consequently, the $4k_F$ Wigner-crystal-like contribution becomes asymptotically dominant over the Friedel contribution. Thus, the condition $g<1/3$ identifies the strongly interacting regime in which the $4k_F$ correlations exhibit a slower spatial decay than the $2k_F$ Friedel contribution.

\begin{figure}[h]
\begin{center}
\includegraphics[scale=0.35]{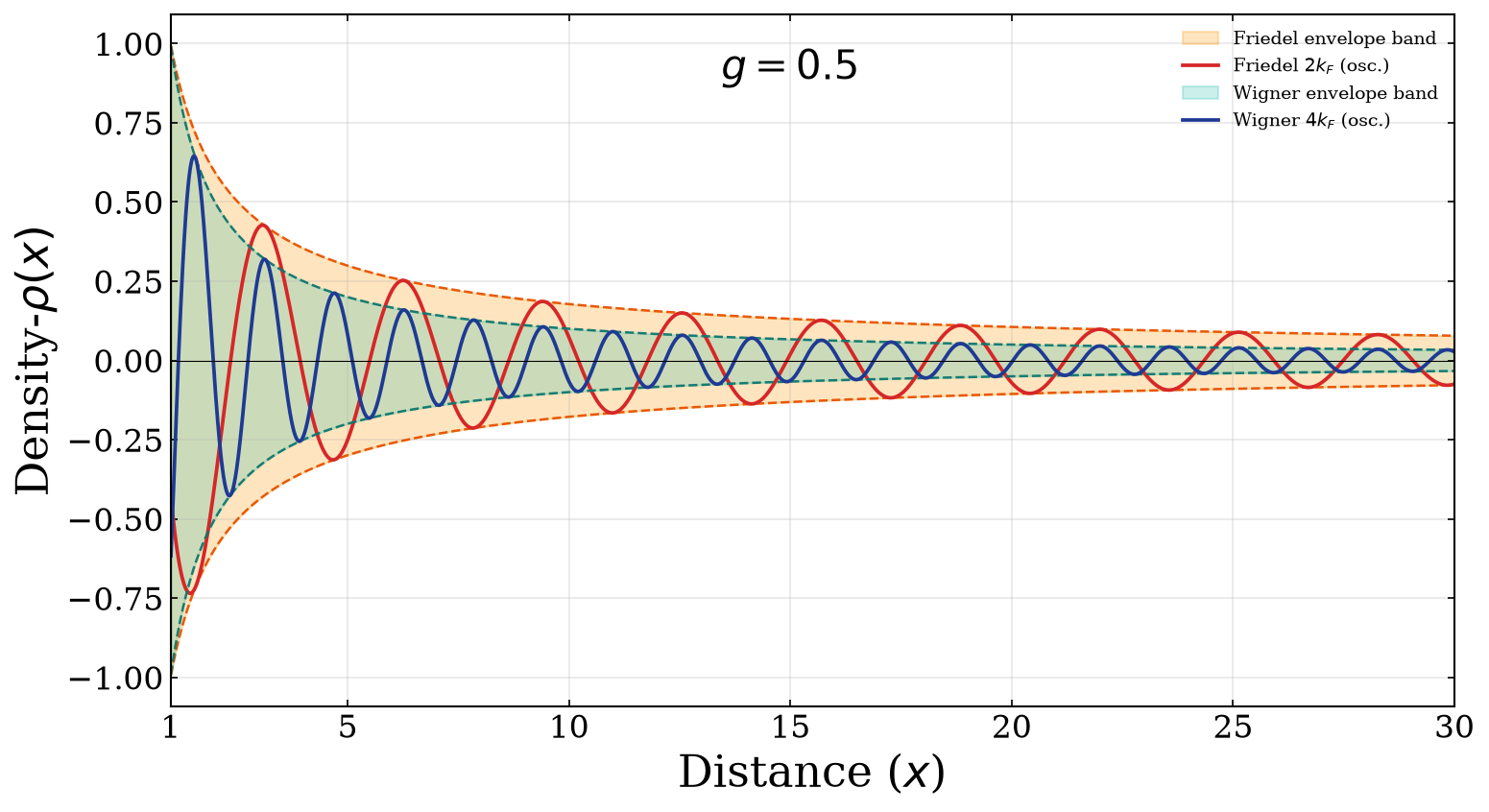}\\
(a)\\
\includegraphics[scale=0.35]{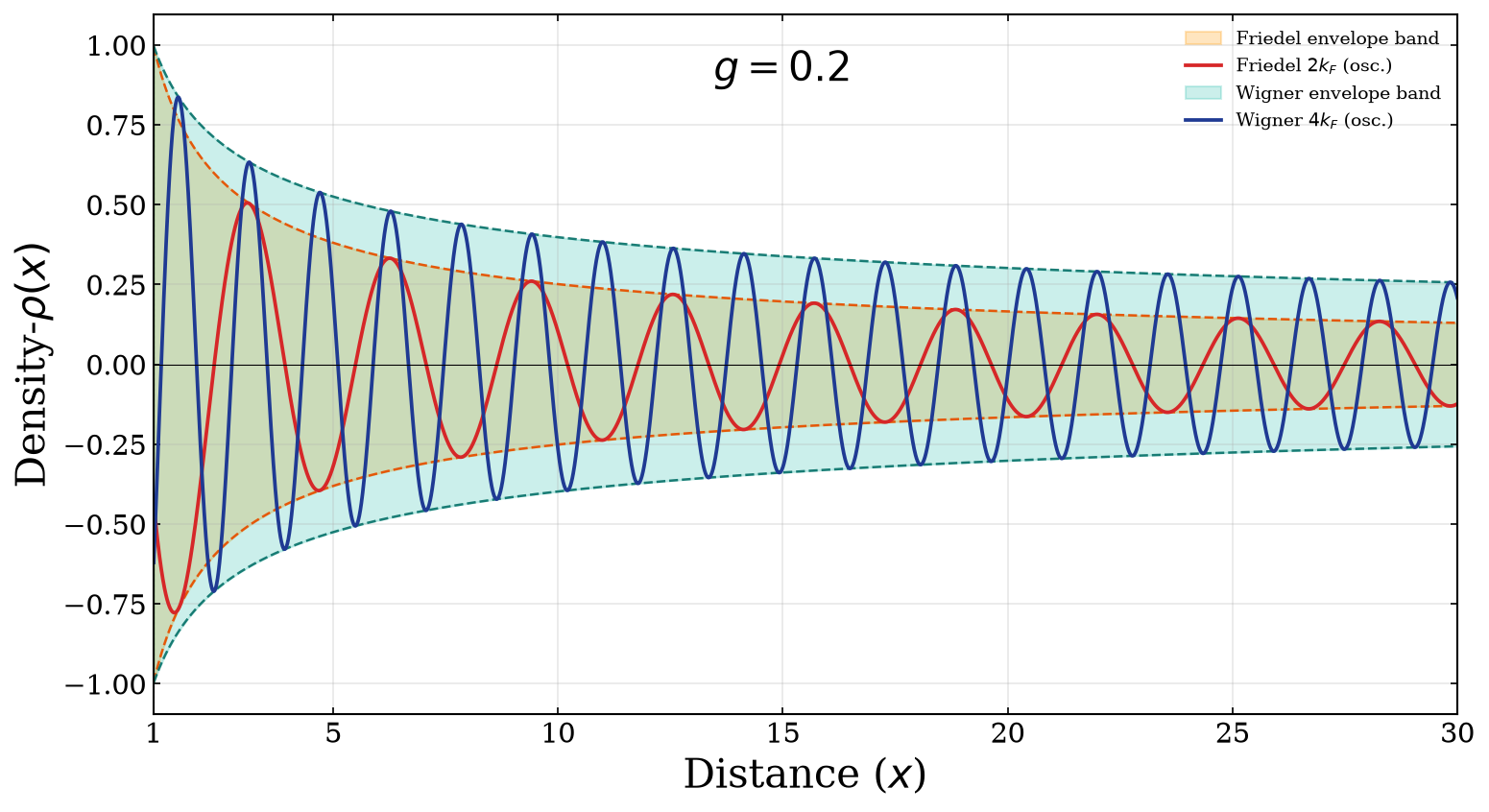}\\
(b)
\end{center}
\caption{Comparison of the $2k_F$ and $4k_F$ density contributions for (a) $g=0.5$ showing the dominance of the Friedel component and (b) $g=0.2$ showing the dominance of the $4k_F$ component.}
\label{FriedelWignerComparison}
\end{figure}
\begin{figure}[h!]
\begin{center}
\includegraphics[scale=0.35]{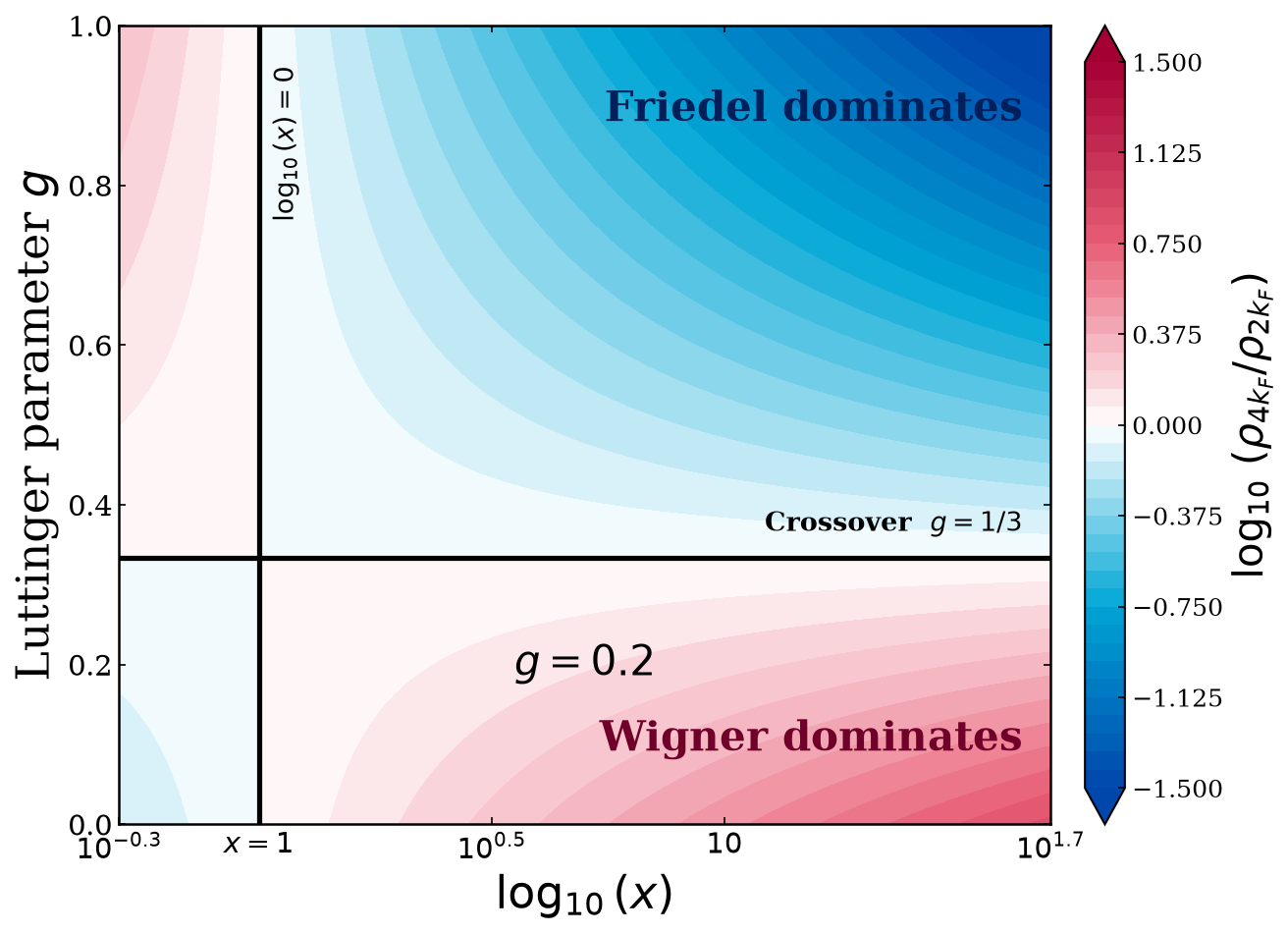}
\end{center}
\caption{Ratio $<\rho_{4k_F}(x)>/<\rho_{2k_F}(x)>$ showing the crossover between $2k_F$ and $4k_F$ dominated regimes at $g=1/3$.}
\label{Ratio}
\end{figure}

The Figures (\ref{FriedelWignerComparison}a) and (\ref{FriedelWignerComparison}b) compare the asymptotic Friedel and \(4k_F\) components at \(g=0.5\) and \(g=0.2\), respectively. For \(g=0.5>1/3\), the \(2k_F\) Friedel contribution decays more slowly and therefore dominates at sufficiently large distance, whereas for \(g=0.2<1/3\), the \(4k_F\) contribution has the slower decay and becomes asymptotically dominant. The Figure (\ref{Ratio}) shows the spatial and interaction dependence of the ratio \(<\rho_{4k_F}(x)>/<\rho_{2k_F}(x)>\) on a logarithmic scale. The resulting phase-space map clearly separates the Friedel-dominated and \(4k_F\)-dominated regions, with the asymptotic crossover at \(g=1/3\). The vertical line at \(x=1\) marks the boundary of the large-distance regime considered here.

\subsection{Reflection-coupled harmonic analysis}

The conventional harmonic analysis of the fermionic field operator provides a convenient representation of the density operator in terms of its slowly varying and rapidly oscillating components. In this representation, the density may be written as 
\small
\begin{equation}
\begin{aligned}
\rho(x,t) = \rho_0 + \tilde{\rho}_s(x,t) + \rho_0 (e^{2ik_Fx} e^{2\pi i\int_{-\infty}^{x}dy\tilde{\rho}_s(y,t)}+h.c.)
\end{aligned}
\label{anomalousscaling}
\end{equation}
\normalsize
where $\tilde{\rho}_s(x,t)$ denotes the slowly varying part of the density fluctuation ($\tilde{\rho}_s= \rho_s-<\rho_s>$). The corresponding rapidly varying contributions can therefore be identified as
\begin{equation}
\begin{aligned}
\rho_f(x,t)
&=\rho_0 e^{2\pi i\int_{-\infty}^{x}dy\,\tilde{\rho}_s(y,t)},\\
\label{haldaneharmonic}
\end{aligned}
\end{equation}
Consequently, the component of the density-density correlation function carrying the oscillatory factor $e^{2ik_F(x-x')}$ takes the form
\small
\begin{equation}
\begin{aligned}
\langle \rho_f(x,t)&\rho_f(x',t')\rangle
\sim
\left\langle
e^{2\pi i\int_{-\infty}^{x}dy\,\tilde{\rho}_s(y,t)}
e^{2\pi i\int_{-\infty}^{x'}dy'\,\tilde{\rho}_s(y',t')}
\right\rangle .
\end{aligned}
\label{HHA}
\end{equation}
\normalsize
In the absence of mutual interactions, the left-hand side of equation (\ref{HHA}) can also be evaluated directly using Wick's theorem in terms of the fermionic two-point functions. For a homogeneous system, the exponential average appearing on the right-hand side of equation (\ref{HHA}) can be evaluated using the Baker--Campbell--Hausdorff (BCH) relation and the corresponding bosonic two-point correlations. The result is then consistent with that obtained independently from the fermionic representation through Wick's theorem. Thus, the conventional harmonic decomposition provides a consistent representation of the rapidly varying density in the translationally invariant case.

For an inhomogeneous system containing a localized impurity, however, this correspondence no longer persists. The exponential average on the right-hand side of equation (\ref{HHA}), when evaluated using the density correlations modified by the impurity, produces non-trivial scaling exponents even in the absence of mutual interactions. In contrast, the direct fermionic calculation based on Wick's theorem continues to yield the trivial exponents appropriate to a non-interacting system. The appearance of such non-trivial exponents in a system in which the fermions do not mutually interact indicates that the conventional harmonic decomposition does not fully capture the reflection-induced structure of the rapidly varying density component in the presence of spatial inhomogeneity.

This observation motivates a non-standard reflection-coupled harmonic analysis in which the harmonic decomposition incorporates the left-right mixing induced by impurity backscattering. Within UFBT, this is achieved by incorporating the spatially reflected density fluctuation into the rapidly varying component. Accordingly, equation (\ref{haldaneharmonic}) is generalized as
\begin{equation}
\begin{aligned}
\rho_f(x,t)
&=\rho_0\,e^{2\pi i\int_{-\infty}^{x}dy\,
\frac{\tilde{\rho}_s(y,t)+\tilde{\rho}_s(-y,t)}{\sqrt{2}}}.
\end{aligned}
\label{ufbtharmonic}
\end{equation}
Here, the $\tilde{\rho}_s(-y,t)$ term incorporates the spatial mixing associated with impurity-induced backscattering, while the factor $1/\sqrt{2}$ provides the normalization of the symmetrized density field. The essential modification is therefore not merely a change in the harmonic amplitude, but a change in the density fields entering the harmonic decomposition. Remarkably, when this reflection-coupled prescription is used, the resulting correlation functions acquire the corresponding trivial exponent structure, and the bosonized result reproduces the result obtained directly from Wick's theorem. This agreement provides a non-trivial consistency check on the reflection-coupled harmonic representation. The reflection-coupled prescription thus forms the basis of the harmonic analysis employed in UFBT.

When equation (\ref{ufbtharmonic}) is employed to evaluate the envelope of the rapidly oscillating component of the density-density correlation function, $\langle\rho_f(x_1,t_1)\rho_f(x_2,t_2)\rangle$, the resulting expression coincides with that obtained independently from the general four-point correlation function in equation (\ref{rhofrhof}). In this comparison, the second cumulant involving the density fields provides the leading singular contribution, while higher connected cumulants represent less singular contributions. The resulting leading term has the power-law scaling that determines the corresponding correlation exponent. Further evaluation at coincident spatial and temporal coordinates, with the appropriate point-splitting prescription and spin degrees of freedom taken into account, yields the scaling exponents for both the impurity-induced Friedel oscillations in equation (\ref{FriedelExp}) and the Wigner-crystal correlations in equation (\ref{wigner3}).

Thus, the reflection-coupled harmonic representation consistently incorporates the impurity-induced left-right mixing and reproduces the correlation functions obtained from the underlying fermionic formulation. Importantly, the agreement is not restricted to the impurity-induced $2k_F$ Friedel sector but extends to the higher-harmonic correlation sector. The resulting structure provides a consistent framework within UFBT for describing the harmonic organization of correlations in strongly inhomogeneous Luttinger liquids.

\subsection{Spinless Luttinger Liquid results}
The conversion from the spinful to the spinless case is straightforward. The Green functions obtained using UFBT for the class of systems considered here can be reduced to the spinless case by eliminating the spin sector and appropriately rescaling the holon exponents. Specifically, the holon exponents are doubled, $P\rightarrow2P$, $Q\rightarrow2Q$ and $X\rightarrow2X$, while all spinon exponents are set to zero. Consequently, the spinless system is characterized by a single renormalized velocity,
\begin{equation}
v_h=\sqrt{v_F^2+\frac{v_Fv_0}{\pi}},
\end{equation}
reflecting the absence of spin-charge separation.

With these substitutions, the Friedel oscillation term in equation (\ref{FriedelExp}) reduces to
\begin{equation}
\begin{aligned}
\langle\rho_f(x)\rangle
&\sim x^{4X-2P-2Q}
\sim x^{-g},
\end{aligned}
\label{FriedelExp2}
\end{equation}
where $g$ denotes the corresponding spinless scaling exponent. This result is consistent with the established result in the literature \cite{egger1995friedel}.\\

\section{Conclusion}
An analytical framework for strongly inhomogeneous one-dimensional Luttinger liquids has been established using the Unified Field Bosonization Technique (UFBT), within which localized impurities are incorporated exactly at the free-fermion level and electron-electron interactions are treated non-perturbatively. Closed analytical expressions for the Green functions and density-density correlations reveal a unified structure containing both impurity-induced $2k_F$ Friedel oscillations and $4k_F$ Wigner-crystal like correlations. Their asymptotic scaling laws, $x^{-(g+1)/2}$ and $x^{-2g}$, respectively, identify a distinct crossover at $g=1/3$: the Friedel contribution dominates for $g>1/3$, whereas sufficiently strong repulsive interactions with $g<1/3$ drive the system into a regime dominated by $4k_F$ Wigner-crystal correlations.

A non-standard reflection coupled harmonic analysis has been formulated to incorporate the left-right mixing generated by impurity backscattering. The resulting harmonic representation reproduces the density correlations obtained independently from the underlying fermionic formulation and consistently yields the scaling exponents of both oscillatory contributions. The spinless limit follows directly by eliminating the spin sector and rescaling the holon exponents. Taken together, these results establish UFBT as a unified analytical framework capable of capturing the interplay between impurity-induced inhomogeneity and interaction-driven correlations beyond the conventional harmonic description of Luttinger liquids.\\

\section*{Acknowledgments}

The author gratefully acknowledges Prof. Girish S. Setlur for his guidance and valuable discussions during the development of the ideas underlying this work.


\bibliographystyle{apsrev4-1}
\bibliography{ref}
\normalsize

\end{document}